\documentstyle[12pt]{article}

\def\hybrid{\topmargin -20pt	\oddsidemargin 0pt
	\headheight 0pt	\headsep 0pt
	\textwidth 6.25in	
	\textheight 9.5in	
	\marginparwidth .875in
	\parskip 5pt plus 1pt	\jot = 1.5ex}

\hybrid

\def\baselinestretch{1.2}

\catcode`\@=11

\def\marginnote#1{}
\newcount\hour
\newcount\minute
\newtoks\amorpm
\hour=\time\divide\hour by60
\minute=\time{\multiply\hour by60 \global\advance\minute by-\hour}
\edef\standardtime{{\ifnum\hour<12 \global\amorpm={am}%
	\else\global\amorpm={pm}\advance\hour by-12 \fi
	\ifnum\hour=0 \hour=12 \fi
	\number\hour:\ifnum\minute<10 0\fi\number\minute\the\amorpm}}
\edef\militarytime{\number\hour:\ifnum\minute<10 0\fi\number\minute}

\def\draftlabel#1{{\@bsphack\if@filesw {\let\thepage\relax
   \xdef\@gtempa{\write\@auxout{\string
      \newlabel{#1}{{\@currentlabel}{\thepage}}}}}\@gtempa
   \if@nobreak \ifvmode\nobreak\fi\fi\fi\@esphack}
	\gdef\@eqnlabel{#1}}
\def\@eqnlabel{}
\def\@vacuum{}
\def\draftmarginnote#1{\marginpar{\raggedright\scriptsize\tt#1}}

\def\draft{\oddsidemargin -.5truein
	\def\@oddfoot{\sl preliminary draft \hfil
	\rm\thepage\hfil\sl\today\quad\militarytime}
	\let\@evenfoot\@oddfoot	\overfullrule 3pt
	\let\label=\draftlabel
	\let\marginnote=\draftmarginnote
   \def\@eqnnum{(\theequation)\rlap{\kern\marginparsep\tt\@eqnlabel}%
\global\let\@eqnlabel\@vacuum}  }

\def\preprint{\twocolumn\sloppy\flushbottom\parindent 2em
	\leftmargini 2em\leftmarginv .5em\leftmarginvi .5em
	\oddsidemargin -.5in	\evensidemargin -.5in
	\columnsep .4in	\footheight 0pt
	\textwidth 10.in	\topmargin  -.4in
	\headheight 12pt \topskip .4in
	\textheight 6.9in \footskip 0pt
	\def\@oddhead{\thepage\hfil\addtocounter{page}{1}\thepage}
	\let\@evenhead\@oddhead	\def\@oddfoot{}	\def\@evenfoot{} }

\def\numberbysection{\@addtoreset{equation}{section}
	\def\theequation{\thesection.\arabic{equation}}}

\def\underline#1{\relax\ifmmode\@@underline#1\else
	$\@@underline{\hbox{#1}}$\relax\fi}

\def\titlepage{\@restonecolfalse\if@twocolumn\@restonecoltrue\onecolumn
     \else \newpage \fi \thispagestyle{empty}\c@page\z@
	\def\thefootnote{\fnsymbol{footnote}} }

\def\endtitlepage{\if@restonecol\twocolumn \else \newpage \fi
	\def\thefootnote{\arabic{footnote}}
	\setcounter{footnote}{0}}  

\catcode`@=12
\relax

\def\figcap{\section*{Figure Captions\markboth
	{FIGURECAPTIONS}{FIGURECAPTIONS}}\list
	{Figure \arabic{enumi}:\hfill}{\settowidth\labelwidth{Figure
999:}
	\leftmargin\labelwidth
	\advance\leftmargin\labelsep\usecounter{enumi}}}
 \relax
\def\tablecap{\section*{Table Captions\markboth
	{TABLECAPTIONS}{TABLECAPTIONS}}\list
	{Table \arabic{enumi}:\hfill}{\settowidth\labelwidth{Table
999:}
	\leftmargin\labelwidth
	\advance\leftmargin\labelsep\usecounter{enumi}}}
 \relax
\def\reflist{\section*{References\markboth
	{REFLIST}{REFLIST}}\list
	{[\arabic{enumi}]\hfill}{\settowidth\labelwidth{[999]}
	\leftmargin\labelwidth
	\advance\leftmargin\labelsep\usecounter{enumi}}}
 \relax
\makeatletter
\newcounter{pubctr}
\def\publist{\@ifnextchar[{\@publist}{\@@publist}}
\def\@publist[#1]{\list
	{[\arabic{pubctr}]\hfill}{\settowidth\labelwidth{[999]}
	\leftmargin\labelwidth
	\advance\leftmargin\labelsep
	\@nmbrlisttrue\def\@listctr{pubctr}
	\setcounter{pubctr}{#1}\addtocounter{pubctr}{-1}}}
\def\@@publist{\list
	{[\arabic{pubctr}]\hfill}{\settowidth\labelwidth{[999]}
	\leftmargin\labelwidth
	\advance\leftmargin\labelsep
	\@nmbrlisttrue\def\@listctr{pubctr}}}
 \relax
\makeatother
\newskip\humongous \humongous=0pt plus 1000pt minus 1000pt

\newif\ifdtup

\relax

\def\thefootnote{\fnsymbol{footnote}}
\def\be{\begin{equation}}
\def\ee{\end{equation}}
\def\ba{\begin{eqnarray}}
\def\ea{\end{eqnarray}}

\begin{document}
\renewcommand{\theequation}{\thesection.\arabic{equation}}
\newcommand{\beq}{\begin{equation}}
\newcommand{\eeq}[1]{\label{#1}\end{equation}}
\newcommand{\ber}{\begin{eqnarray}}
\newcommand{\eer}[1]{\label{#1}\end{eqnarray}}
\begin{titlepage}
\begin{center}

\hfill CERN-TH/95-258\\
\hfill FTUAM/95-34\\
\hfill hep-th/9510028\\

\vskip 1.0in

{\large \bf SUPERSYMMETRY AND DUALITIES}
\footnote{Contribution to the proceedings of the Trieste conference
on
{\em S-Duality and Mirror Symmetry}, 5--9 June 1995; to be published
in
Nuclear Physics B, Proceedings Supplement Section, edited by E. Gava,
K. Narain and C. Vafa}

\vskip 0.7in

{\bf Enrique Alvarez
\footnote{Permanent address: Departamento de Fisica Teorica,
Universidad
Autonoma, 28049 Madrid, Spain}, Luis Alvarez-Gaume and Ioannis Bakas}
\footnote{Permanent address:  Department of Physics, University of
Patras,
26110 Patras, Greece}\\
\vskip .1in

{\em Theory Division, CERN\\
     CH-1211 Geneva 23, Switzerland}\\

\vskip .1in

\end{center}

\vskip .8in

\begin{center} {\bf ABSTRACT } \end{center}
\begin{quotation}\noindent
Duality transformations with respect to rotational isometries relate
supersymmetric with non-supersymmetric backgrounds in string theory.
We find that non-local world-sheet effects have to be taken into
account in order to restore supersymmetry at the string level. The
underlying superconformal algebra remains the same, but in this case
T-duality relates local with non-local realizations of the algebra in
terms of parafermions. This is another example where stringy effects
resolve paradoxes of the effective field theory.
\end{quotation}
\vskip1.7cm
CERN-TH/95-258 \\
FTUAM/95-34\\
October 1995\\
\end{titlepage}
\vfill
\eject

\def\baselinestretch{1.2}
\baselineskip 16 pt
\section{\bf Outline and summary}
\setcounter{equation}{0}
\noindent
These notes, based on two lectures given at the Trieste conference on
``S-Duality and Mirror Symmetry", summarize some of our recent
results
on supersymmetry and duality [1, 2, 3]. The problem that arises in
this context
concerns the supersymmetric properties of the lowest order effective
theory and their behaviour under T-duality
(and more generally U-duality) transformations (see also [4]).
We will describe the geometric conditions for the Killing vector
fields
that lead to the preservation or the loss of manifest space-time
supersymmetry under duality, and classify them as translational
versus
rotational respectively. Analyzing this problem in detail we find
that
there is no
contradiction at the level of string theory. This is part of the
standard lore that stringy effects manifest as paradoxes of the
effective field theory, and supersymmetry is no exception to it.

In toroidal compactifications of superstring theory T-duality is a
symmetry that interchanges momentum with winding modes, and it
manifests geometrically as a small-large radius equivalence in the
effective theory of the background massless fields. This equivalence
has
been generalized to arbitrary string backgrounds with Abelian
isometries
following Buscher's formula that was originally derived for the
$\beta$-function equations of the metric $G_{\mu \nu}$, antisymmetric
tensor
field $B_{\mu \nu}$ and dilaton $\Phi$ to lowest order in the
$\sigma$-model
coupling constant (inverse string tension) ${\alpha}^{\prime}$ (see
for
instance [5], and references therein). One might
naively expect that if a background is supersymmetric (eg a bosonic
solution
of effective supergravity), the dual background will also share the
same
number of supersymmetries. This is a reasonable expectation provided
that
there are no non-local world-sheet effects, so that
the dual faces of the theory provide a trustworthy low energy
approximation
to string dynamics in the ${\alpha}^{\prime}$-expansion.
There are circumstances, however, when
the dual geometry has strong curvature singularities and the
effective
field theory description breaks down in their vicinity.
This is a typical property of
gravitational backgrounds having Killing isometries with fixed
points, in which
case the T-dual background exhibits curvature singularities. Then,
T-duality
appears to be incompatible with space-time supersymmetry, and
non-local
world-sheet effects have to be taken into account for resolving this
issue
consistently. Afterall, it is a well-known fact that symmetries not
commuting with a given Killing vector field are not manifest in the
dual
background of the effective field theory description.

In the first part most of our discussion will concentrate
on 4-dim space-time backgrounds, thinking of superstring vacua as
arising from ten dimensions by compactification on a 6-dim internal
space
$K$. It is much simpler to expose the main ideas in this case and
employ the notion of self-duality (when it is appropriate) to refine
the distinction we want to make. In four dimensions, and in the
presence
of some Killing isometries, it is also possible
to intertwine T with S-duality, thus forming a much larger symmetry
group
known as U-duality. We will present the essential features of this
idea
at the level of the effective theory by
considering 4-dim backgrounds with (at least)
one isometry, in which case the U-duality corresponds to $O(2,2)$
transformations. The supersymmetric properties of the effective
theory
will also be considered under the action of such generalized
symmetries,
where the effect on supersymmetry can be even more severe.

In the second part we will investigate more elaborately the issue of
supersymmetry versus duality in heterotic $\sigma$-models, where we
analyze possible anomalies and find some modifications of Buscher's
rules.
We consider the heterotic string in 10-dim flat Minkowski space with
$SO(32)$ gauge group (though similar arguments apply to the
$E_8 \times E_8$ string) and study the effect of T-duality with
respect
to rotations in a 2-dim plane, using a manifest
$ISO(1, 7) \times SO(30) \times SO(2)$ symmetric formalism.
We prove that the number
of space-time fermionic symmetries remains unchanged. The conformal
field theory
analysis suggests that the emission vertices of low energy particles
in
the dual theory are represented by non-local operators, which do not
admit a straightforward ${\alpha}^{\prime}$-expansion. In fact the
underlying
conformal theory provides a natural explanation of this, where local
realizations of the superconformal algebra become non-local in terms
of
parafermions. In view of these circumstances, certain theorems
relating
world-sheet with manifest space-time supersymmetry should be revised.
Also, the powerful constraints imposed by supersymmetry on the
geometry of the
target space are not valid for non-local realizations.

Some of the topics of this work may also be relevant for superstring
phenomenology, where we mainly work in terms of the lowest order
effective theory. The issue of space-time supersymmetry versus
duality,
which arises to lowest order in ${\alpha}^{\prime}$, demonstrates
explicitly that an apparently non-supersymmetric background can
qualify
as a vacuum solution of superstring theory, in contrary to the
``standard wisdom" that has been considered so far. So, whether
supersymmetry is broken or not in various phenomenological
applications
cannot be decided, unless one knows how to incorporate the
appropriate non-local world-sheet effects that might lead to its
restoration
at the string level. Also, various gravitational solutions, like
black holes,
might enjoy some supersymmetric properties in the string context.
This
could also provide a better understanding of the way that string
theory, through its world-sheet effects, can resolve the fundamental
problems of the quantum theory of black holes. Finally, from the low
energy point of view the fact that T-duality relates supersymmetric
with
non-supersymmetric backgrounds could provide examples of the
mechanism
advocated by Witten to shed new light in the cosmological constant
problem [6].
Hence, we view T-duality at this moment as a method for probing some
of these
possibilities, and leave to future investigation the applicability of
these ideas to the physical problems of black holes and cosmology
within
superstring theory.

\section{\bf Supersymmetry and dualities: a first clash}
\setcounter{equation}{0}
\noindent
Consider the class of 4-dim backgrounds with
one isometry generated by a Killing vector field
$K = \partial / \partial \tau$. The signature is taken to be
Euclidean, but most of the results generalize to the
physical Minkowskian signature.
The metric can be written locally in the form
\be
ds^2 = V (d \tau + {\omega}_{i} dx^{i})^2 + V^{-1} {\gamma}_{ij}
dx^{i} dx^{j} ~ ,
\ee
where $x^{1} = x$, $x^{2} = y$, $x^{3} = z$ are coordinates on the
space of
non-trivial orbits of $\partial / \partial \tau$, and $V$,
${\omega}_{i}$,
${\gamma}_{ij}$ are all independent of $\tau$. Of course,
${\omega}_{i}$ are
not unique, since they are defined up to a gauge transformation
${\omega}_{i} \rightarrow {\omega}_{i} - {\partial}_{i} \lambda$ that
amounts
to the coordinate shift $\tau \rightarrow \tau + \lambda (x^{i})$.
These
backgrounds provide gravitational solutions of the $\beta$-function
equations
with constant dilaton and antisymmetric tensor fields to lowest order
in
${\alpha}^{\prime}$. Although this class is rather restrictive, it is
a good strating point for exploring the clash of supersymmetry with
dualities
in the effective theory.

Performing a T-duality transformation to the gravitational background
(2.1)
we obtain the following configuration in the $\sigma$-model frame,
\be
G_{\mu \nu} = {1 \over V} \left(\begin{array}{ccccccc}
1 & & 0 & & 0 & & 0 \\
0 & &   & &   & &  \\
0 & & &  & {\gamma}_{ij} &  &  \\
0 & &  &  &  &  &  \end{array} \right)
\ee
with non-trivial dilaton and antisymmetric tensor fields
\be
\Phi = -{1 \over 2} \log V ~ , ~~~~~ B_{\tau i} = {\omega}_{i} ~ .
\ee
A necessary condition for having manifest space-time supersymmetry in
the new
background is given by the dilatino variation $\delta \lambda = 0$,
\be
{1 \over 2} H_{\mu \nu \rho} \pm \sqrt{\det G} ~
{{\epsilon}_{\mu \nu \rho}}^{\sigma} {\partial}_{\sigma} \Phi = 0 ~ ,
\ee
where $H_{\mu \nu \rho}$ is the field strength of $B_{\mu \nu}$.
Substituting
the expressions (2.2) and (2.3) we arrive at the differential
equations
\be
{\partial}_{i} {\omega}_{j} - {\partial}_{j} {\omega}_{i} \pm
{\sqrt{\det \gamma} \over 2 V^3} {{\epsilon}_{\tau i j}}^{k}
{\partial}_{k} V
= 0 ~ ,
\ee
which select only a subclass of solutions (2.1) of the vacuum
Einstein equations.

It is useful at this point to introduce a non-local quantity $b$,
\be
{\partial}_{i} b = {1 \over 2} V^2 \sqrt{\det \gamma} ~
{{\epsilon}_{i}}^{jk}
({\partial}_{j} {\omega}_{k} - {\partial}_{k} {\omega}_{j}) ~ ,
\ee
where ${\epsilon}_{ijk}$ is the fully antisymmetric tensor with
respect to the
metric $\gamma$ with ${\epsilon}_{123} =1$. It is then clear that the
dilatino variation of the dual background is zero provided that $b
\pm V$
is constant. This constraint can be reformulated as an axionic
instanton
condition on the dual background, because $V = e^{-2 \Phi}$, and $b$
coincides with the axion field associated with $B_{\tau i} =
{\omega}_{i}$.
Another useful interpretation of $b$ in the initial frame (2.1) is
given
by the nut potential of the original metric [7]. We note that $b$
exists
only on-shell. Furthermore, the identification of the nut potential
of the
original metric with the axion field of the T-dual background is
important
for describing not only the necessary condition for having manifest
space-time supersymmetry, but also later for understanding the
concept of
U-duality at work.

The original background being purely gravitational trivially
satisfies the
dilatino equation. Manifest space-time supersymmetry requires the
existence
of a Killing spinor $\eta$, which in turns implies the existence of a
complex structure as a bilinear form in $\eta$ and $\bar{\eta}$ in
the usual way.
But a Kahlerian Ricci-flat metric is also hyper-Kahler, and hence
self-dual
(or anti-self-dual),
\be
R_{\mu \nu \rho \sigma} = \pm {1 \over 2} \sqrt{\det G} ~
{{\epsilon}_{\rho \sigma}}^{\kappa \lambda} R_{\mu \nu \kappa
\lambda} ~ .
\ee
Consequently, exploring the behaviour of space-time supersymmetry for
T-duality transformations of purely gravitational backgrounds forces
us to
consider self-dual solutions of the lowest order effective theory. In
this
case the original backgrounds exhibit $N=4$ extended superconformal
symmetry associated with the three underlying complex structures.
According to the standard rules of manifest supersymmetry it is
natural
to expect that the dual background (2.2) and (2.3) will also exhibit
$N=4$ world-sheet supersymmetry, and its metric will be conformally
equivalent to a hyper-Kahler metric $G_{\mu \nu}^{\prime}$ (in the
presence of non-trivial $B_{\mu \nu}$ and $\Phi$) so that
\be
G = \Omega G^{\prime} ~ , ~~~~~ {\Box}^{\prime} \Omega = 0 ~ .
\ee
It will be demonstrated later that this is possible only for $b \pm V
= 0$
(up to a constant). Otherwise manifest space-time supersymmetry is
lost,
and non-local realizations of the $N=4$ superconformal algebra have
to be
introduced for resolving this paradox of the effective theory at the
string level. In this case the dual background admits no Killing
spinors,
thus providing a vacuum solution of superstring theory that is
apparently non-supersymmetric.

A potential trouble with supersymmetry has already been spotted in
the
adapted coordinate system (2.1) for the Killing isometry
$\partial / \partial \tau$.
The covariant description of
the criterion for having manifest space-time supersymmetry after
duality
can be formulated as follows: we divide the Killing vector fields
$K_{\mu}$
into two classes, the translational and the rotational. The first
class
consists of Killing vector fields with self-dual (respectively
anti-self-dual)
covariant derivatives
\be
{\nabla}_{\mu} K_{\nu} = \pm {1 \over 2} \sqrt{\det G} ~
{{\epsilon}_{\mu \nu}}^{\rho \sigma} {\nabla}_{\rho} K_{\sigma} ~ ,
\ee
while the second class encompasses all the rest. Consider now the
conjugate
fields $S_{\pm} = b \pm V$ and introduce the quadratic quantity
\be
\Delta S_{\pm} = {\gamma}^{ij} {\partial}_{i} S_{\pm} {\partial}_{j}
S_{\pm} ~ ,
\ee
which is clearly $\geq 0$. It is a well-known theorem that $\Delta
S_{\pm} = 0$
for translational isometries (in which case $S_{\pm}$ is constant),
while
$\Delta S_{\pm} > 0$ for rotational isometries [8]. In the latter
case it is
always possible to choose the coordinates $x$, $y$ and $z$ so that
$S_{\pm} = z$. This analysis provides the covariant distinction
between those
isometries that preserve manifest space-time supersymmetry after
duality
and those that do not. The result can be easily extended from purely
gravitational backgrounds to more
general solutions with $N=4$ superconformal symmetry having
antisymmetric
tensor and dilaton fields. Indeed, if the original background has a
metric
which is hyper-Kahler up to a conformal factor, as in (2.8), the
T-dual background
will also be manifestly supersymmetric if the Killing vector
field is translational with
respect to $G^{\prime}$.

Next, we investigate in detail the specific form of hyper-Kahler
metrics with
one isometry, and use it to determine the dependence of the
corresponding
complex structures and the Killing spinors on the Killing coordinate
$\tau$.
It turns out that for rotational isometries these quantities depend
explicitly
on $\tau$, which is the key for having non-local realizations of
supersymmetry
after T-duality. This problem does not arise for translational
isometries
as nothing depends explicitly on $\tau$; for this reason
translational
isometries are also known as tri-holomorphic.

\underline{Translational isometries} : In this case the adapted
coordinate
system (2.1) can be chosen so that
\be
{\gamma}_{ij} = {\delta}_{ij} ~ , ~~~~
{\partial}_{i}(V^{-1}) = \pm {1 \over 2} {\epsilon}_{ijk}
({\partial}_{j} {\omega}_{k} - {\partial}_{k} {\omega}_{j}) ~ ,
\ee
and so self-dual metrics with translational isometry are determined
by
solutions of the 3-dim Laplace equation
\be
({\partial}_{x}^2 + {\partial}_{y}^2 + {\partial}_{z}^2) V^{-1} = 0 ~
{}.
\ee
The localized solutions are of the general form [7, 9]
\be
V^{-1} = \epsilon + \sum_{i=1}^{n} {m_{i} \over \mid \vec{x} -
\vec{x_{0}}_{i} \mid} ~ ,
\ee
where $\epsilon$ is a constant that determines the asymptotic
behaviour of
the metric and $m_{i}$, $\vec{x_{0}}_{i}$ are moduli parameters. The
apparent singularities of the metric are removable if all $m_{i} = M$
and
$\tau$ is taken to be periodic with range $0 \leq \tau \leq 4 \pi
M/n$.
For $\epsilon = 0$ the resulting metrics are the multi-center
Gibbons-Hawking
metrics, with $n=2$ being the simplest non-trivial example known as
the
Eguchi-Hanson instanton. For $\epsilon \neq 0$ (in which case its
value is
normalized to 1) one has the multi-Taub-NUT metrics, with $n=1$ being
the
ordinary self-dual Taub-NUT metric.

The three independent complex structures are known to be
$\tau$-independent.
Choosing ${\omega}_{3} = 0$ for convenience, we have the following
result for
the corresponding Kahler forms in the special frame (2.11) [10]:
\ba
F_{1} & = & (d \tau + {\omega}_{2} dy) \wedge dx - V^{-1}
dy \wedge dz ~ , \nonumber \\
F_{2} & = & (d \tau + {\omega}_{1} dx) \wedge dy + V^{-1} dx \wedge
dz ~ , \\
F_{3} & = & (d \tau + {\omega}_{1} dx + {\omega}_{2} dy) \wedge dz -
V^{-1} dx \wedge dy ~ . \nonumber
\ea
As for the Killing spinors of the background (2.1), (2.11) one
can easily check that they are the constant, independent of any
coordinates.

\underline{Rotational isometries} : In this case the self-duality
condition
reduces to a non-linear equation in three dimensions involving a
function $\Psi (x, y, z)$. The metric (2.1) can always be chosen
so that [8]
\ba
{\omega}_{1} & = & \mp {\partial}_{y} \Psi ~ , ~~~~
{\omega}_{2} = \pm {\partial}_{x} \Psi ~ , ~~~~ {\omega}_{3} = 0 ~ ,
\nonumber\\
V^{-1} & = & {\partial}_{z} \Psi ~ , ~~~~~~
{\gamma}_{ij} = diag ~ (e^{\Psi}, ~ e^{\Psi}, ~ 1) ~ ,
\ea
where $\Psi$ satisfies the continual Toda equation
\be
({\partial}_{x}^2 + {\partial}_{y}^2) \Psi + {\partial}_{z}^2
e^{\Psi} = 0 ~ .
\ee
It can be verified directly that $\Delta S_{\pm} = 1$, and hence
$S_{\pm} = z$
indicating the anomalous behaviour of manifest space-time
supersymmetry
under rotational T-duality transformations.

The Kahler forms that describe the three independent complex
structures in
this case are not all $\tau$-independent; one of them is a
$SO(2)$-singlet,
while the other two form a $SO(2)$-doublet. We have explicitly the
following result [2] for the doublet,
\be
\left(\begin{array}{c}
F_{1} \\
F_{2}   \end{array} \right) = e^{{1 \over 2} \Psi}
\left(\begin{array}{ccr}
\cos {\tau \over 2} &  & \sin {\tau \over 2} \\
\sin {\tau \over 2} & & - \cos {\tau \over 2} \end{array} \right)
\left(\begin{array}{c}
f_{1} \\
f_{2}  \end{array}  \right) ,
\ee
where
\ba
f_{1} & = & (d \tau + {\omega}_{2} dy) \wedge dx -
V^{-1} dz \wedge dy ~ , \nonumber\\
f_{2} & = & (d \tau + {\omega}_{1} dx) \wedge dy + V^{-1} dz \wedge
dx
\ea
and for the singlet,
\be
F_{3} = (d \tau + {\omega}_{1} dx + {\omega}_{2} dy) \wedge dz +
V^{-1} e^{\Psi} dx \wedge dy ~ .
\ee
It is straightforward to verify that they are covariantly constant
on-shell and
satisfy the $SU(2)$ Clifford algebra, as required.
As for the Killing spinors in this case
one finds that they are constant spinors up to an overall phase
$e^{\pm i \tau /4}$ that depends explicitly on $\tau$.

Rotational isometries are more rare than translational isometries in
4-dim
hyper-Kahler geometry. The only example known to this date with only
rotational isometries is the Atiyah-Hitchin metric, whereas other
metrics like
the Eguchi-Hanson and Taub-NUT exhibit both (see for instance [10]).
The
flat space also exhibits both
type of isometries. Choosing $\Psi = \log z$ as the simplest solution
of the
continual Toda equation we obtain the metric
\be
ds^2 = z d{\tau}^2 + dx^2 + dy^2 + {1 \over z} dz^2 ~ ,
\ee
which is the flat space metric written in terms of the coordinates
$2 \sqrt{z} \cos (\tau / 2)$ and $2 \sqrt{z} \sin (\tau / 2)$.
Introducing
$r^2 = 4z$ and $\theta = \tau / 2$ we obtain the metric in standard
polar coordinates,
\be
ds^2 = dx^2 + dy^2 + dr^2 + r^2 d{\theta}^2 ~ .
\ee
This example is rather instructive because it provides a good
approximation of
target space metrics around the fixed points (located at $r = 0$) of
a
Killing isometry. It is clear that the dual background will have a
curvature singularity at $r=0$, where the $\sigma$-model formulation
of strings
is expected to break down. If one insists on exploring the dual
metric from the
low energy point of view by determining the corresponding dilaton
field to
lowest order in ${\alpha}^{\prime}$, space-time supersymmetry will
appear to
be lost. Clearly more powerful techniques should be used in order to
understand the behaviour of strings close to this point.

One might think that any isometry with fixed points could lead to
anomalous
behaviour of space-time supersymmetry under duality,
as it is seen from the effective theory viewpoint. This is however
not true
in general. The relevant analysis of this issue is rather simple for
4-dim
spaces $M$. At a fixed point $p$ the action of $K$ gives rise to an
isometry
$T_{p}(M) \rightarrow T_{p}(M)$ on the tangent space, which is
generated by
the antisymmetric $4 \times 4$ matrix ${\nabla}_{\mu} K_{\nu}$. Any
such
matrix can have rank 2 or 4, provided that the Killing vector field
$K$ is
not zero everywhere. For rank 2 the fixed points form a 2-dim
subspace
called bolt. The metric close to a bolt can be approximated by (2.21)
and
the apparent singularity at
$r = 0$ is nothing but a coordinate singularity in the flat polar
coordinate
system on $R^2$. For rank 4, $p$ is an isolated fixed point called
nut after
the fixed point at the center of the self-dual Taub-NUT metric. In
this
case the metric has a removable singularity and it is approximated
using the flat polar coordinate system on $R^4$ centered at the nut.
Consider now for general $M$ (not necessarily hyper-Kahler) the
decomposition
\be
{\nabla}_{\mu} K_{\nu} = K_{\mu \nu}^{+} + K_{\mu \nu}^{-}
\ee
into
self-dual and anti-self-dual parts. A nut is said to be self-dual (or
anti-self-dual) if $K_{\mu \nu}^{\pm}$ is zero (ie., $K$ is
translational)
at $p$. If $M$ is hyper-Kahler, as it is the case of interest here,
then
$K$ will be translational everywhere. Therefore we can have a Killing
isometry
with a fixed point (self-dual nut) which preserves manifest
space-time
supersymmetry under T-duality. For bolts, however, we always have
\be
K_{\mu \nu}^{+} K^{+ \mu \nu} = K_{\mu \nu}^{-} K^{- \mu \nu} ~ ,
\ee
and so $K_{\mu}$ cannot be translational since otherwise both
$K_{\mu \nu}^{\pm} = 0$ (ie., $K_{\mu} $ will be constant everywhere
if $M$ is
hyper-Kahler). Only in this case manifest supersymmetry
behaves anomalously under duality. For this reason we
may consider the bolt-type metric (2.21) as a characteristic example
of rotational isometries. In the next section we will study the
10-dim
analogue of this in heterotic string theory and perform duality
transformations
with respect to rotations in a 2-dim plane.

We return now to the general situation where the T-duality
transformation is
performed on a purely gravitational background (2.1). The standard
description of T-duality as a canonical transformation in $\tau$ and
its
conjugate momentum [11] amounts to
\be
\tau \rightarrow \int (V^{-1} \partial \tau - {\omega}_{1} \partial x
- {\omega}_{2} \partial y) dz - (V^{-1} \bar{\partial} \tau +
{\omega}_{1} \bar{\partial} x + {\omega}_{2} \bar{\partial} y) d
\bar{z} ~ ,
\ee
where we have set ${\omega}_{3} = 0$. Applying this formula to the
frame with translational isometries we find that the complex
structures
(2.14) become in the dual background [2]
\be
{\tilde{F}}_{i}^{\pm} = V^{-1} (\pm d \tau \wedge dx^{i} -
{1 \over 2} {\epsilon}_{ijk} dx^{j} \wedge dx^{k}) ~ .
\ee
Here $\pm$ refers to the right or left structures, which are not the
same because
the duality generates non-trivial torsion (2.3). We also note in this
case
that the dual metric (2.2) is conformally flat (and hence
hyper-Kahler) with
a conformal factor $\Omega = V^{-1}$ that satisfies the Laplace
equation
(2.12) in
agreement with (2.8). Hence, the dual background exhibits $N=4$
superconformal
symmetry which is locally realized with the aid of the dual complex
structures,
and space-time supersymmetry is manifest.

For rotational isometries the duality transformation (2.24) acts
differently on the
complex structures and yields the dual forms
\be
{\tilde{F}}_{3}^{\pm} = V^{-1} (\pm d \tau \wedge dz + e^{\Psi}
dx \wedge dy) ~ ,
\ee
and
\ba
{\tilde{f}}_{1}^{\pm} & = & V^{-1} (\pm d \tau \wedge dx -
dz \wedge dy) ~ , \nonumber\\
{\tilde{f}}_{2}^{\pm} & = & V^{-1} (\pm d \tau \wedge dy +
dz \wedge dx) ~ .
\ea
Then, we see explicitly that ${\tilde{F}}_{1,2}^{\pm}$ become
non-locally
realized in terms of the target space fields [2]. These non-local
variables
can be used explicitly to construct a new realization of the $N=4$
superconformal
algebra, which is preserved under duality. Indeed only the
realization
changes form, since otherwise T-duality would not be a string
symmetry.
Hence, we conclude that non-local world-sheet effects have to be
taken into
account in order to restore the space-time supersymmetry that is
apparently
lost in this case. If one takes seriously the low energy effective
theory
will face the paradox that the dual metric (2.2) is not conformally
hyper-Kahler anymore. In fact, even if we have one local complex
structure
(provided in this example by ${\tilde{F}}_{3}$) we do not seem to
have manifest
$N=1$ space-time supersymmetry,
although the converse is always true. Further details on the
resolution
of this issue will be presented in the next section.

The non-local realization of the $N = 4$ superconformal
algebra that arises in this context is reminiscent of the
parafermionic
realizations in conformal field theory. Unfortunately there is no
exact
conformal field theory description of the backgrounds (2.2), (2.3)
and
so to strengthen this analogy we appeal to another example that a
similar problem arises after T-duality. Namely, consider the pair of
coset models $SU(2) \times U(1)$ and $(SU(2)/U(1)) \times U(1)^{2}$,
which
are related by T-duality with respect to a rotational isometry. Both
models have $N=4$ world-sheet supersymmetry, but in the latter it is
non-locally realized in terms of the $SU(2)/U(1)$ parafermions [12].
The
relevant analysis for the transformation of the underlying complex
structures has been performed in this case [2], in exact analogy with
the above
discussion, but we omit the details here. The background that arises
to lowest order in ${\alpha}^{\prime}$ for the second coset turns out
not
to be manifestly space-time supersymmetric.

A note of general value is
that Killing spinors with explicit dependence on the Killing
coordinate
$\tau$ are not maintained after duality.

Finally, we discuss briefly the behaviour of supersymmetry under
U-duality
transformations. For 4-dim string backgrounds with one isometry it is
possible to intertwine T with S-duality non-trivially and produce new
symmetry generators. The key point in this investigation is provided
by the Ehlers transform, which for pure gravity is a continuous
$SL(2)$
symmetry acting on the space of vacuum solutions [13, 7].
It acts non-locally, as its
formulation requires the notion of the nut potential $b$. More
precisely,
the dimensional reduction of 4-dim gravity for the metrics (2.1)
reads
as follows,
\be
\int d^4 x \sqrt{\det G} ~ R^{(4)} \rightarrow \int d^3 x \sqrt{\det
\gamma}
\left( R^{(3)} - {1 \over 2} V^{-2} ({\partial}_{i} V {\partial}^{i}
V -
{\partial}_{i} b {\partial}^{i} b) \right) ,
\ee
and so $b \pm V$ form a conjugate pair of $SL(2) / U(1)$
$\sigma$-model
variables. The celebrated Ehlers transform is
\ba
V^{\prime} & = & {V \over (Cb + D)^2 - C^2 V^2} ~ , \nonumber\\
b^{\prime} & = & {(AD + BC)b + AC(b^2 - V^2) + BD \over (Cb + D)^2 -
C^2 V^2} ~ ,
\ea
where $AD - BC = 1$. Recall now the observation that the nut
potential
coincides with the axion field of the T-dual background (2.2), (2.3).
Therefore the Ehlers transform behaves as an S-duality
transformation, where
one starts from a purely gravitational solution and reaches the new
one
(2.29) via a sequence of T-S-T transformation within the context of
the
string effective theory (switching on and then off again non-trivial
torsion
and dilaton fields). Straightforward generalization of this argument
to
the full massless sector leads to an enlarged symmetry of the
$\beta$-function
equations that is called U-duality (given in its
continuous form). For 4-dim backgrounds
S and T-S-T are two $SL(2)$ symmetries that combine into an $O(2, 2)$
group [1],
while for the 10-dim heterotic string compactified on a 7-dim torus
this
procedure yields the bigger group $O(8, 24)$ [14].

U-duality transformations have a more severe effect on
supersymmetries,
provided that the intertwining of S with T is performed using
rotational
isometries. It is rather instructive for this purpose to consider the
class
of self-dual metrics and ask whether the Ehlers transform always
preserves the
self-duality. The answer is yes for translational isometries and no
for
rotational. A simple example to demonstrate this is provided by the
$SL(2)$ action (2.29) on the flat space metric (2.20) written in
polar coordinates.
Choosing $A = D = 1$
and $B = 0$ we may verify that the new metric fails to be self-dual
[1].
On the contrary, the same group element acts in the translational
frame by
a simple shift
\be
V^{-1} \rightarrow V^{-1} + 2C ~ ,
\ee
and while it preserves the self-duality it has a non-trivial effect
on the
boundary conditions; starting from (2.13) with $\epsilon = 0$,
the parameter $C$ generates
solutions with $\epsilon \neq 0$. The main point we wish to make here
is that
rotational isometries, from the
space-time point of view in the T-S-T dual face, apparently destroy
all three
complex structures; otherwise Ricci-flatness would imply self-duality
for the transformed metric (2.29).

It will be interesting to explore the possibility to have non-local
realizations of supersymmetry in this case
by reformulating the Ehlers symmetry as a non-local transformation on
the
target space coordinates. This way we hope to extend the results of
our
investigation on ``supersymmetry versus duality" to the most general
situation for U-duality symmetries that arise by compactification to
three or even two
space-time dimensions [15].

\section{\bf Duality in heterotic string theory}
\setcounter{equation}{0}
\noindent
We begin this section by working out the T-duality transformation for
a general $(1, 0)$ heterotic $\sigma$-model with arbitrary connection
and background gauge field. We find that if one does not want to have
a non-local dual world-sheet action, due to anomalies which appear
when
implementing the duality transformation, one has to transform under
the
isometry the right-moving fermions. This yields a non-trivial
transformation
of the background gauge field under T-duality [3]. We also find that
if in the
original model the gauge and the spin connections match, so that
there is
anomaly cancellation in the effective theory, the change in the gauge
field
under T-duality ensures the same matching condition
in the dual theory. Furthermore,
if the original theory had $(2, 0)$ or $(2, 2)$ superconformal
invariance,
the dual theory also has these properties.

We consider a manifold $M$ with metric $G_{ij}$, antisymmetric tensor
field $B_{ij}$ and a background gauge connection $V_{iAB}$ associated
to
a gauge group $G \in O(32)$ (for simplicity we consider the $O(32)$
heterotic
string). We introduce the $(1, 0)$ superfields
\be
X^{i} (\sigma, \theta) = x^i + \theta {\lambda}^i ~ , ~~~~
{\Psi}^A (\sigma, \theta) = {\psi}^{A} + \theta F^A ~ ,
\ee
where $x^i (\sigma)$ are the fields embedding the world-sheet
in the target space, $F^A$ are auxiliary fields, and the fermions
$\lambda$
and $\psi$ have opposite world-sheet chirality. Using light-cone
coordinates on the world-sheet,
${\sigma}^{\pm} = ({\sigma}^{0} \pm {\sigma}^{1}) / \sqrt{2}$,  and
defining the operator
\be
D = {\partial \over \partial \theta} + i \theta
{\partial \over \partial {\sigma}^{+}} ~ , ~~~~ D^2 = i
{\partial}_{+} ~ ,
\ee
and
\be
{\cal D} \Psi^A = D\Psi^A + V_{i}\,^{A}\,_{B}(X) DX^i \Psi^B ~ ,
\ee
we consider the Lagrangian density
\be
L = \int d\theta \left(-i(G_{ij} + B_{ij}) DX^i
\partial_{-}X^j
- \delta_{AB} \Psi^A {\cal D} \Psi^B \right) ~ .
\ee
Eliminating the auxiliary fields one obtains [16]
\be
L = (G_{ij} + B_{ij}) \partial_{+} x^i \partial_{-} x^j
+ i G_{ij} \lambda^i D_{-} \lambda^j + i \psi^A  D_{+} \psi^A +
{1 \over 2} F_{ijAB} \lambda^i \lambda^j \psi^A \psi^B ~ ,
\ee
with world-sheet supercurrent of type (1,0),
\be
{\cal G}_{+} = (2 G_{ij} + B_{ij})\partial_{+} x^i \lambda^j
- {i \over 2} H_{ijk} \lambda^i \lambda^j \lambda^k ~ .
\ee

We will carry out a duality transformation with respect to an
isometry
of the metric that leaves (3.4) and (3.5) invariant.
Following the procedure outlined in [17],
we gauge the isometry, with
some gauge fields $A_{\pm}$, and add an extra term with a Lagrange
multiplier
making the gauge superfield strength vanish. If we integrate out the
Lagrange multiplier, the gauge superfields become pure gauge.
Using the invariance of the action we can change variables to remove
all
presence of gauge fields and recover the original action. If instead
we
integrate first over $A_{\pm}$
and then fix the gauge, we obtain the dual theory. In
our case we will perform these steps in a manifestly
$(1,0)$-invariant
formalism. Recall that the necessary condition for gauging an
isometry generated by a Killing vector field $K^{i}$ of the metric is
[18]
\be
K^i H_{ijk} = \partial_j U_k - \partial_k U_j ~ ,
\ee
and
\be
\delta_{K} B_{ij} = \partial_i (K^l B_{lj} + U_j) -
\partial_j (K^l B_{li}) ~ ,
\ee
for some vector $U$. Then, the
conserved (1,0)-supercurrent for the first term of (3.4) is
\be
{\cal J_{-}} = (K_i - U_i) \partial_{-} X^i, ~~~~
{\cal J_{+}} = (K_i + U_i) DX^i ~ ,
\ee
so that
$ D {\cal J_{-}} + \partial_{-} {\cal J_{+}} = 0 .$

We now introduce $(1, 0)$ gauge fields ${\cal A_{-}} = A_{-} + \theta
\chi_{-}$
and ${\cal A} = \chi + i \theta A_{+}$ of bosonic
and fermionic character respectively.
If $\epsilon (\sigma,\theta)$ is the gauge parameter, we can take
$\delta_{\epsilon}{\cal A_{-}} = - \partial_{-} \epsilon$ and
$\delta_{\epsilon}{\cal A} = - D \epsilon$, and with the variation
$\delta_{\epsilon} X^i = \epsilon K^i(X)$ the
gauge invariant Lagrangian (assuming that $K^i U_i$ is constant)
is
\be
(G_{ij} + B_{ij}) DX^i \partial_{-} X^j + {\cal J_{+}}{\cal A_{-}}
+ {\cal J_{-}}{\cal A} + K^2 {\cal A_{-}}{\cal A} ~ .
\ee
The left-moving part
$\Psi (D \Psi + V_i DX^i)\Psi$
is invariant under this global transformation when the isometry
variation can be compensated by a gauge transformation
\ba
\delta X^i & = & \epsilon K^i (X) ~ , ~~~~ \delta \Psi = - \kappa
\Psi ~ , \nonumber\\
\delta_K V_i & = & {\cal D}_i \kappa = \partial_{i} \kappa
+[V_i,\kappa] ~ ,
\ea
which implies
\be
K^i F_{ij} = D_j \mu ~ ;  ~~~~~
\mu = \kappa - K^i V_i ~ .
\ee
Making $\epsilon$ a function of $(1,0)$ superspace one obtains after
some
algebra
$\delta_{\epsilon} ( \Psi^{T} {\cal D} \Psi) = D\epsilon \Psi^{T} \mu
\Psi$,
and hence, adding the coupling
$ {\cal A} \Psi^{T}\mu\Psi $
we achieve gauge invariance.
The full gauge invariant Lagrangian reads
\be
L = -i \left( (G_{ij} + B_{ij}) DX^i \partial_{-} X^j + {\cal
J_{+}}
{\cal A_{-}} + {\cal J_{-}} {\cal A} + K^2 {\cal A_{-}} {\cal A}
\right)
- (\Psi^{T} {\cal D} \Psi + {\cal A} \Psi^{T} \mu \Psi) ~ .
\ee

Add now the Lagrange multiplier superfield term
$i \Lambda (D {\cal A_{-}} - \partial {\cal A})$
and integrate out ${\cal A}$ and ${\cal A_{-}}$ to obtain the dual
Lagrangian
\be
{\tilde L}_{cl} = -i \left( ({\tilde G}_{ij} + {\tilde B}_{ij}) DX^i
\partial_{-}X^j + ({\cal J}_{+} + D \Lambda ){1\over K^2}
(\partial_{-}
\Lambda
+ i \Psi^{T} \mu \Psi - {\cal J}_{-}) \right)
- \Psi^{T} {\cal D} \Psi ~ .
\ee
In coordinates adapted to the Killing vector
the dual values for ${\tilde G}_{ij}$, ${\tilde B}_{ij}$, ${\tilde
V}_{i}$ are
\ba
{\tilde G}_{00} & = & {1\over K^2} ~ , ~~~~
{\tilde G}_{0 \alpha} = {1\over K^2} U_{\alpha} ~ , ~~~~~
{\tilde G}_{\alpha\beta} = G_{\alpha\beta} - {K_{\alpha} K_{\beta} -
U_{\alpha}U_{\beta} \over K^2} ~ , \nonumber\\
{\tilde B}_{0\alpha} & = & - {1\over K^2} K_{\alpha} ~ , ~~~~
{\tilde B}_{\alpha\beta} = B_{\alpha\beta} +{K_{\alpha} B_{0\beta} -
K_{\beta} B_{0\alpha} \over K^2} ~ ,\\
{\tilde V}_{0 AB} & = & - {1\over K^2} \mu_{AB} ~ , ~~~~
{\tilde V}_{\alpha AB} = V_{\alpha AB} - {1\over K^2} (K_{\alpha} +
U_{\alpha}) \mu_{AB} ~ . \nonumber
\ea
These results are equivalent to
Buscher's formulae (see for instance [5]),
but in this case we find a change in
the background gauge field as well.

The preceding formulae were obtained using only classical
manipulations.
In general, however, there will be anomalies and the dual action may
not
have the same properties as the original one. Depending on the choice
for $\mu$
and the gauge group $G \in O(32)$, the dual theory may be afflicted
with
anomalies, in which case the two theories are not equivalent.
Equivalence would follow provided we include some Wess-Zumino-Witten
terms
generated by the
quantum measure.
If we want the two local Lagrangians $L$ and ${\tilde L}$ to be
equivalent, we must
find the conditions on $G$, $B$, $V$ and $\mu$ in order to cancel the
anomalies.
Using standard techniques in anomaly computations, the variation of
the
effective action is expressible in the form (see [3] for details)
\ba
\delta \Gamma_{eff} & = & -{1\over 4\pi} \int Tr (\omega_i
\partial_{-}
X^i - {\cal A}_{-} \Omega)D(\epsilon (K^j \omega_j + \Omega)) d^3 Z
\nonumber\\
& & -{1 \over 4 \pi} \int Tr(V_i D X^i -{\cal A} \Omega) \partial_{-}
(\epsilon (K^j V_j + \mu)) d^3 Z ~ ,
\ea
where $d^3 Z = d^2 \sigma d \theta$, and we have
introduced the effective gauge field
\be
V_{-}\,^a\,_b = \omega_{-}\,^a\,_b - A_{-} \Omega^a\,_b  ~ .
\ee

Note that the dual theory will contain
non-local contributions unless we cancel the anomaly,
which is a mixture between $U(1)$
and $\sigma$-model
anomalies [19].
The simplest way to cancel it is to assume that the spin and
gauge connection match in the original theory, a condition that also
makes
the space-time anomalies to cancel (see for instance [20]).
In this case if $\mu =
\Omega$
with matching quadratic Casimirs, the anomalous variation
$\delta {\Gamma}_{eff}$
can be cancelled by a local counterterm.
A test of the validity of the generalized duality transformation is
that if
we start with a theory with matching spin and gauge connection
$\omega = V$,
the dual theory is also guaranteed to have ${\tilde \omega} = {\tilde
V}$.
It is straightforward to verify this condition in our case.
This is also important for the consistency of the model with respect
to
global world-sheet and target-space anomalies
and it implies that if the original theory is conformally invariant
to
$O(\alpha^{\prime})$, so is the dual theory.

There is yet one more possible source of anomalies under duality if
the
original
model is $(2,0)$ or $(2,2)$-superconformal invariant. In the $(2,2)$
case for
instance we have a $U(1)_{L} \times U(1)_{R}$ current algebra.
The manifold has a covariantly constant complex structure,
$\nabla_k J^i\,_j = 0$, \quad $ J^i\,_k J^k\,_j = - \delta^i\,_j$.
The
R-symmetry
is generated by the rotation $\delta \lambda^i = \epsilon J^i\,_j
\lambda^j$
with current ${\cal J}_{+} = i J_{ij} \lambda^i \lambda^j$.
This current has an anomaly
\be
\partial_{-} {\cal J}_{+} = -{1\over 4\pi} R_{ijk}\,^l J_l\,^k
\partial_{+}x^i \partial_{-} x^j  ~ ,
\ee
which can be removed only if the right-hand side is
cohomologically trivial. From [17] we know that T-duality preserves
$N=2$
global supersymmetry,
hence, we should be able to improve the dual R-current so that the
$U(1)_{L}
\times U(1)_{R}$ current algebra is preserved, as needed for the
application of
the
theorem in [21]. Since generically a T-duality transformation
generates a
non-constant dilaton, the energy-momentum tensor of the dual theory
contains
an improvement term due to the dilaton $\Phi$ of the form
$\partial_{+}^2 \Phi$.
As a consequence of $N=2$ global supersymmetry there should also be
an
improvement term in the fermionic currents and in the $U(1)$
currents. Since
the one-loop $\beta$-function implies (in complex coordinates)
$R_{\alpha{\bar \beta}} \sim \partial_{\alpha}\partial_{{\bar \beta}}
\Phi$,
we can improve
the $U(1)_{L} \times U(1)_{R}$ currents so that they are chirally
conserved. In the (2,2) case the improvements are
\ba
\Delta{\cal J}_{+} & = & \partial_{\alpha} \Phi \partial_{+}
Z^{\alpha} - \partial_{{\bar \alpha}} \Phi \partial_{+} Z^{{\bar
\alpha}}
{}~ , \nonumber\\
\Delta{\cal J}_{-} & = & -(\partial_{\alpha} \Phi \partial_{-}
Z^{\alpha}
- \partial_{{\bar \alpha}} \Phi \partial_{-} Z^{{\bar \alpha}}) ~ .
\ea
With these improvements the currents are chirally conserved to order
$\alpha^{\prime}$ (and presumably to all orders, since the higher
loop
counterterms are cohomologically trivial for a $(2,2)$ supersymmetric
$\sigma$-model); hence we conclude that under duality the $(2,0)$ or
$(2,2)$
superconformal algebra is preserved, as was also pointed
out earlier. We meet the conditions
to apply the Banks et al theorem [21] implying that the theory is
space-time
supersymmetric.

In the following we give a concrete answer to the puzzles raised in
section 2 for space-time supersymmetry in the context of 10-dim
heterotic string theory. We consider the motion in flat Minkowski
space
\be
ds^2 = dr^2 + r^2 d{\theta}^2 + (dx^{i})^2 - (dx^{0})^2 ~ , ~~~~
i = 1, \cdots , 7
\ee
and revisit the problem of ``space-time supersymmetry versus duality"
first from the effective action point of view, and then within the
framework
of conformal field theory. The frame (3.20) provides the natural
generalization
of the bolt-type coordinates (2.21). Before duality we have
the full $ISO(1,9)$ Lorentz invariance and $O(32)$
gauge symmetry.
Since we perform duality in (3.20) with respect to rotations in a
2-dim
plane, only the subgroup of $ISO(1,9)$ commuting with them will be a
manifest
local symmetry of the effective action. Similarly if we preserve
manifest (1,0)
supersymmetry on the world-sheet and avoid anomalies, we embed the
isometry
group $SO(2) \subset G \equiv SO(32)  $. The subgroup of $G$
commuting with
$SO(2)$ is
$SO(30) \times SO(2)$ and once again this will be a manifest symmetry
in the low energy theory. It is well known that
under T-duality, symmetries not commuting with the ones generating
duality
are generally realized non-locally. Hence although the dual
background
still contains all the original symmetries from the CFT point of
view,
the low energy theory does not seem to exhibit them. The theory will
be
explicitly symmetric under $ISO(1,7) \times SO(30) \times SO(2)$
only.
We want to make sure nevertheless that the original space-time
supersymmetry
is preserved, although not in a manifest $O(1,9)$-covariant
formalism.
For this we can consider the variation
of the fermionic degrees of freedom in a formalism adapted to the
$ISO(1,7)
\times SO(30) \times SO(2)$ symmetry, and look for which combination
of
the $O(1,9)$ fermions are annihilated by supersymmetry.

The low energy approximation to the heterotic string is given by
$N=1$ supergravity coupled to $N=1$ super Yang-Mills in $d=10$.
In ten dimensions
we can impose simultaneously the Majorana and Weyl conditions [22];
then,
in terms of $SO(1,7)$,
a Majorana-Weyl spinor of $SO(1,9)$ becomes a Weyl spinor. Write the
Dirac
algebra (in an orthonormal frame) as
\ba
\Gamma_{\mu} & = & \tau_3 \otimes \gamma_{\mu}\ ; ~~~
\mu = 0,1,...,7 ~ , \nonumber\\
\Gamma_{7+i} & = & i \tau_i \otimes 1\ ; ~~~ i =1, 2 ~ ,\\
{\bar \Gamma} & = & \tau_3 \otimes \gamma_9 ~ , \nonumber
\ea
where $\gamma_9$ is the analogous of the 4-dim $\gamma_5$
in eight dimensions. Ten dimensional
indices will be hatted.
The supersymmetric variation of the
ten-dimensional fermions is given by
\ba
\delta{\hat \Psi}_{{\hat \mu}} & = &
(\partial_{{\hat \mu}} - {1 \over 4}
\omega_{{\hat \mu}{\hat a}{\hat b}} \Gamma^{{\hat a}{\hat b}})
{\hat \epsilon} ~ , \nonumber\\
\delta {\hat \lambda} & = & (\Gamma^{{\hat \mu}} \partial_{{\hat
\mu}} \Phi -
{1\over 6} H_{{\hat \mu}{\hat \nu}{\hat \rho}} \Gamma^{{\hat
\mu}{\hat \nu}
{\hat \rho}}){\hat \epsilon} ~ , \\
\delta \chi^A & = & -{1\over 4} F^A\,_{{\hat \mu}{\hat \nu}}
\Gamma^{{\hat \mu}{\hat \nu}}{\hat \epsilon} \nonumber
\ea
for the gravitino, dilatino and gluino, respectively.

The dual background is
\ba
d{\tilde s}^2 & = & dr^2 +{1\over r^2} d{\tilde \theta}^2
-
(dx^0)^2 +(dx^i)^2 \ ; ~~~ i = 1,...,7 , \nonumber\\
\Phi & = & - \log r ~ , ~~~~~
V_{\mu} dx^{\mu} = {1\over r^2} d{\tilde \theta} M ~ ,
\ea
where $M$ is the matrix describing the embedding of the spin
connection
in the gauge group, which we take to be the standard one acting only
on two
of the right-moving fermions.
The background gauge
field
strength is
\be
F = -{2\over r^3} dr \wedge d{\tilde \theta} ~ ,
\ee
and so decomposing the above variations with respect to $SO(1,7)$ we
find:
for the gravitino
\ba
\delta \Psi_{\mu} & = & \partial_{\mu} \epsilon ~ , ~~~~
\delta \Psi_{\{r\}} = \partial_{\{r\}}\epsilon ~ , \nonumber\\
\delta \Psi_{\{{\tilde \theta}\}} & = & (\partial_{\theta} + {i \over
4 r^2}
\tau_3 \otimes 1)\epsilon ~ ,
\ea
for the dilatino
\be
\delta \lambda = -{ i\over r} (\tau_1 \otimes 1)  \epsilon ~ ,
\ee
and for the gluino
\be
\delta \chi^A = 0 ~ ,  ~~~~
\delta \chi =- {i\over r^2} (\tau_3 \otimes 1)
\epsilon
\ee
for $A \in SO(32)$ and along the embedded $SO(2)$ respectively.
In the preceding formulas $\epsilon$ is an $SO(1,7)$ Weyl
spinor with the same number of independent components as a 10-dim
Majorana-Weyl spinor.

If we define now
\ba
{\tilde \Psi}_{\mu} & = & \Psi_{\mu} ~ , ~~~~
{\tilde \Psi}_{\{r\}} = \Psi_{\{r\}} ~ , \nonumber\\
{\tilde \Psi}_{\{{\tilde \theta}\}} & = & \Psi_{\{{\tilde \theta}\}}
+ {i \over 4}e^{\Phi}  (\tau_2 \otimes 1) \lambda ~ ,
\ea
it is easy to see that the new fields transform as
$\delta {\tilde
\Psi}_{{\hat \mu}} = \partial_{{\hat \mu}} \epsilon$. Similarly,
\ba
{\tilde \lambda} & = & \lambda  + i e^{-\Phi} (\tau_2 \otimes 1)
\chi ~, \nonumber\\
{\tilde \chi} & = & \chi - i e^{\Phi} (\tau_2 \otimes 1) \lambda
\ea
have vanishing variation under space-time supersymmetry. Furthermore,
they
have the correct chiralities as dictated by the ten-dimensional
multiplet.
Thus if we use a formalism covariant only under the explicit $SO(1,7)
\times
SO(30) \times SO(2)$ symmetry of the dual background
we recover the full number of supersymmetric charges. From the
low-energy
effective action this is the most we could expect since at the level
of
the world-sheet CFT the full symmetry $SO(1,9) \times SO(32)$ is only
realized
non-locally. If we want to consider the complete symmetry and the
complete
massless spectrum in the dual theory it seems that the only
reasonable thing
to do is to go back to the two-dimensional point of view.

In the remaining part we will see how to
obtain in principle the vertex operators for the full massless
spectrum in
the dual theory by investigating in detail the way the full symmetry
is
realized from the conformal field theory point of view. The main
point is
to show explicitly that there are indeed world-sheet operators in the
dual theory associated to the space-time supersymmetry charges,
although
some world-sheet non-locality is generated. We will find at this end
an
interesting interplay between the picture-changing operator and
T-duality.

We are considering
the effect of rotational duality in a 2-dim plane, and so the
relevant part of the free heterotic Lagrangian is
\be
L = \partial_{+}{\vec x} \cdot \partial_{-}{\vec x} +
i {\vec \lambda} \cdot \partial_{-}{\vec \lambda} +
i \psi^A \partial_{+} \psi^A + \cdots ~ ,
\ee
where the vector quantities are two-dimensional.
The isometry is $\vec{x} \rightarrow \exp (i \alpha {\sigma}_{2})
\vec{x}$,
but for the time being we shall work in an unadapted frame. Hence
for (3.30) we can perform duality only in the bosonic sector.
The world-sheet supercurrent is
\be
{\cal G}_{+} = {\vec \lambda} \cdot \partial_{+} {\vec x} = {\vec
\lambda}
\cdot {\vec P}_{+} ~ ,
\ee
where ${\vec P}_{+}$ is a chiral current generating translations in
the
target space.
It is convenient to work in canonical pictures [23] ($-{1\over 2}$
for fermion vertices, $-1$ for boson vertices).
The space-time supersymmetry charge is
\be
Q_{\alpha}\,^{(-{1\over 2})} = \oint e^{-{\phi \over 2}}
S_{\alpha} ~ ,
\ee
where $\phi$ is the scalar which bosonizes the superconformal ghost
current
and $S_{\alpha}$ is the spin-field associated to the
$\lambda$-fermions.
The translation operator in the $-1$ picture is
\be
P_{\mu}\,^{(-1)} = \oint e^{-\phi} \lambda_{\mu} ~ .
\ee
Note that in these definitions only the space-time fermion and the
$(\beta ,\gamma)$-ghosts appear. Hence
\be
\{Q_{\alpha}\,^{(-{1\over 2})} , Q_{\beta}\,^{(-{1\over 2})} \}
= \Gamma^{\mu}P_{\mu}\,^{(-1)}
\ee
is satisfied, and if we choose to perform duality for the bosonic
part of
the Lagrangian
only, the same relationships should still hold.

{}From this point of view there is clearly no problem with space-time
supersymmetry.
However, in constructing scattering amplitudes we need to use
vertex operators in different pictures. Hence any problem should come
from the interplay with the picture changing operator ${\cal P}$.
The picture changing operator acting on a vertex operator $V_q (z)$
in the
$q$-picture  can be represented as [23]
\be
{\cal P} V_q (z) = lim_{w \rightarrow z} e^{\phi(w)}
{\cal G}_{+}(w) V_q(z) ~ .
\ee
The only possible difficulties may appear in anomalies in the
world-sheet
supercurrent under duality. Since ${\cal G}_{+}$ does not commute
with
purely bosonic rotations, after duality ${\cal G}_{+}$ will become
non-local
in the world-sheet.
To guarantee that there are no problems with ${\cal G}_{+}$ we want
to make
sure
that the dual world-sheet supercurrent still has the form ${\vec
\lambda} \cdot
{\tilde {\vec P}}_{+}$, where ${\tilde P}_{+}$ is the representation
of the
translation current in the dual theory, and it is here that
the non-locality resides. In fact the full theory in (3.30) can be
constructed
out of the knowledge that $P_{+}\,^i$ ($i = 1, 2$) is chirally
conserved
and that its operator product expansion (OPE) is
$P^i (z) P^j (w) \sim {\delta^{ij} \over (z - w)^2}$. It is hard to
believe that the existence of the chiral currents is going to be lost
under duality. To make sure that this is not the case, the simplest
thing to do is to include sources for these currents and then follow
their transformation under duality.

Following [17]
we gauge the symmetry $\vec{x} \rightarrow \exp(i \alpha
{\sigma}_{2}) \vec{x}$
and concentrate only on the bosonic part of
(3.30), the only one relevant due to the previous arguments.
Thus our starting point is
\be
L = D_{+} x^{T}  D_{-} x + \Lambda F_{+-} ~ ,
\ee
where
$D_{\pm} x = \partial_{\pm} x + i {\sigma}_{2} x A_{\pm}$,
$F_{+-} = \partial_{+} A_{-} - \partial_{-}A_{+}$ and $\Lambda$ is a
Lagrange
multiplier. Here and in the following we drop for convenience the
vectorial
notation for $\vec{x}$ and $\vec{\lambda}$, and introduce $\epsilon =
i {\sigma}_{2}$.
Using the $\Lambda$-equation of motion, $A_{\pm} =
\partial_{\pm}\alpha$,
$D_{\pm} x = e^{- \alpha \epsilon} \partial_{\pm}(e^{\alpha
\epsilon}x)$,
and
changing variables $ x \rightarrow e^{-\alpha \epsilon} x$, the
original theory is recovered.
It proves convenient to parametrize locally
\be
A_{+} = \partial_{+} \alpha_L ~ , ~~~~~
A_{-} = \partial_{-} \alpha_R  ~ .
\ee
Then (3.36) has the symmetries
\be
\delta x = e^{-\epsilon \alpha_{R}} a_R ~ , ~~~~
\delta \lambda
= - x^T \epsilon e^{-\epsilon \alpha_R} a_R ~ ,
\ee
and
\be
\delta x = e^{-\epsilon\alpha_L} a_L ~ , ~~~~
\delta \lambda = x^T
\epsilon e^{-\epsilon \alpha_L} a_L  ~ ,
\ee
yielding respectively the conserved currents
\be
\partial_{-}(e^{\epsilon\alpha_R} D_{+} x) = 0 ~ , ~~~~
\partial_{+}(e^{\epsilon \alpha_L} D_{-} x) =0 ~ .
\ee
When $\alpha_R = \alpha_L = \alpha$, we recover the original currents
$\partial_{\pm}(e^{\epsilon\alpha} x)$. Acting on $x$, these
symmetries
commute.

The sources to be added to (3.36) should be gauge invariant,
\be
J_{-} e^{\epsilon\phi_R} D_{+} x + J_{+} e^{\epsilon \phi_L} D_{-} ~
{}.
\ee
The exponents are non-local in $A_{\pm}$, and they make the coupling
gauge invariant. This also guarantees that the currents in (3.41)
satisfy
the OPE of the original theory as expected.
In our example, let us take for simplicity $J_{+} = 0$. The most
straightforward
way to integrate out the gauge fields is to work in the light-cone
gauge
 $A_{-} = 0$. Then the integral over $A_{+}$ becomes a
$\delta$-function
which can be solved in two ways. If we choose to solve it in order to
write
the Lagrange multiplier $\Lambda$ as a function of the other fields,
we
recover the original theory.

On the other hand if we choose to solve the adapted coordinate to the
isometry
in terms of $\Lambda$ and the other variables, we obtain the dual
theory.
Furthermore we also obtain a determinant, which when properly
evaluated
([5] and references therein) yields the transformation of the
dilaton.
Using polar coordinates $r$, $\theta$ for the 2-dim plane
we obtain the equation
\be
\partial_{-} w = w {i\over r^2} \partial_{-} \Lambda
- {r \over 2 i}(J_{-} - J_{-}^{*} w^2) ~ ,
\ee
where
$w = e^{i \theta}$ and $J_{-}^{*}$ is the complex conjugate of
$J_{-}$.
This is a Riccatti equation, which can be solved order by order in
$J_{-}$.
Restoring powers of $\alpha^{\prime}$ we have
\be
\partial_{-}w = i w {\alpha^{\prime}\over r^2}\partial_{-}
\Lambda - {r \over 2 i \alpha^{\prime}}(J_{-} - J_{-}^{*} w^2) ~ .
\ee
The lowest order solution is
\be
w = \exp i \int {\alpha^{\prime} \over r^2}\partial_{-}
\Lambda d\sigma^{-} ~ ,
\ee
and to this order the current looks like
\be
\partial_{+} \left(r e^{i \theta[\Lambda,J_{-}]}
\right) = \partial_{+}\left(
\rho e^{i\alpha^{\prime}\int {1\over r^2}\partial_{-}\Lambda
d \sigma^{-}}\right) .
\ee
The extra terms depending on $J_{-}$, $J_{-}^{*}$ are required
to guarantee the equality between the correlation functions before
and
after the duality transformation.
To leading order the currents are:
\be
\partial_{+} \left( r e^{\pm i \alpha^{\prime}
\int {1\over
r^2} \partial_{-} \Lambda d \sigma^{-}} \right) , ~~~~
\partial_{-} \left(r e^{\pm i \alpha^{\prime} \int {1\over
r^2}\partial_{+}
\Lambda d\sigma^{+}} \right) .
\ee
Note however that in solving (3.43) there will be corrections to all
orders in $\alpha^{\prime}$ in order to obtain the correct OPE's for
the
dual currents ${\tilde P}_{\pm}$ . These currents can be used to
write the emission vertex operators in the dual theory and they are
almost
always non-local. Since the OPE's of ${\tilde P}_{\pm}$ are
preserved, the
spectrum of the original and the dual theories are equivalent.
Nevertheless, we
have to be careful regarding the
operator mapping.

In conclusion, there is no problem with space-time supersymmetry from
the
point of view of CFT, but the correct operators that needed to be
used to
represent the emission vertices of low energy particles in the dual
theory are
often non-local, and do not admit a straightforward
${\alpha}^{\prime}$
expansion unless we write the dual states in terms of those which
follow from the correspondence as dictated by the duality
transformation.
When there are curvature singularities the approach based on the
effective
low energy theory has many limitations and to obtain reliable
information
we should go back to the underlying string theory. Finally to find
the
graviton, gravitino, etc vertex operators in the dual
picture we could have solved the anomalous dimension operators in the
dual
background (including the dilaton and the background gauge fields),
as was
done for tachyons in [24]. We believe that the two approaches are
equivalent.

\vskip2.0cm
\centerline{\bf REFERENCES}
\begin{enumerate}
\item I. Bakas, Phys. Lett. \underline{B343} (1995) 103.
\item I. Bakas and K. Sfetsos, Phys. Lett. \underline{B349} (1995)
448.
\item E. Alvarez, L. Alvarez-Gaume and I. Bakas, ``T-Duality and
Space-Time Supersymmetry", preprint CERN-TH/95-201, FTUAM-95-101,
hep-th/9507112, July 1995.
\item E. Bergshoeff, R. Kallosh and T. Ortin, Phys. Rev.
\underline{D51}
(1995) 3003; S.F. Hassan, ``T-Duality and Non-local Supersymmetries",
preprint CERN-TH/95-98, hep-th/9504148, April 1995.
\item E. Alvarez, L. Alvarez-Gaume and Y. Lozano, Nucl. Phys. (Proc.
Supp.)
\underline{41} (1995) 1; A. Giveon, M. Porrati and E. Rabinovici,
Phys. Rep. \underline{244} (1994) 77.
\item E. Witten, Int. J. Mod. Phys. \underline{A10} (1995) 1247.
\item G. Gibbons and S. Hawking, Comm. Math. Phys. \underline{66}
(1979) 291.
\item C. Boyer and J. Finley, J. Math. Phys. \underline{23} (1982)
1126;
J. Gegenberg and A. Das, Gen. Rel. Grav. \underline{16} (1984) 817.
\item T. Eguchi, P. Gilkey and A. Hanson, Phys. Rep. \underline{66}
(1980) 213.
\item G. Gibbons and P. Rubback, Comm. Math. Phys. \underline{115}
(1988) 267.
\item E. Alvarez, L. Alvarez-Gaume and Y. Lozano, Phys. Lett.
\underline{B336}
(1994) 183.
\item C. Kounnas, Phys. Lett. \underline{B321} (1994) 26.
\item J. Ehlers, in Les Theories Relativistes de la Gravitation,
CNRS,
Paris, 1959.
\item A. Sen, Nucl. Phys. \underline{B434} (1995) 179.
\item I. Bakas, Nucl. Phys. \underline{B428} (1994) 374; A. Sen,
Nucl. Phys.
\underline{B447} (1995) 62.
\item C. Hull and E. Witten, Phys. Lett. \underline{B160} (1985) 398;
R. Brooks, F. Muhammad and S. Gates, Nucl. Phys. \underline{B268}
(1986)
599; S. Randjbar-Daemi, A. Salam and J. Strathdee, Nucl. Phys.
\underline{B320} (1989) 221; G. Moore and P. Nelson, Nucl. Phys.
\underline{B274} (1986) 509.
\item M. Rocek and E. Verlinde, Nucl. Phys. \underline{B373} (1992)
630.
\item C. Hull, Mod. Phys. Lett. \underline{A9} (1994) 161.
\item G. Moore and P. Nelson, Phys. Rev. Lett. \underline{53} (1984)
1519.
\item M. Green, J. Schwarz and E. Witten, ``Superstrings", Cambridge
University Press, 1987.
\item T. Banks, L. Dixon, D. Friedan and E. Martinec, Nucl. Phys.
\underline{B299} (1988) 613; T. Banks and L. Dixon,
Nucl. Phys. \underline{B307} (1988) 93.
\item L. Alvarez-Gaume, ``An Introduction to Anomalies", in Erice
School in
Mathematical Physics, Erice, 1985; P. van Nieuwenhuizen, ``An
Introduction
to Supersymmetry, Supergravity and the Kaluza-Klein Program", in Les
Houches
1983, North Holland, 1984.
\item D. Friedan, E. Martinec and S. Shenker, Nucl. Phys.
\underline{B271}
(1986) 93.
\item C. Callan and Z. Gan, Nucl. Phys. \underline{B272} (1986) 647.
\end{enumerate}

\end{document}